\documentstyle[proc,epsf]{article}

\begin{document}
\begin{center}
{\Large\bf Multiplicity Fluctuations in Au+Au Collisions at RHIC}
\end{center}

\begin{center}
 {\large Krzysztof Wo\'zniak} \\
{\it Institute of Nuclear Physics PAN,
ul. Radzikowskiego 152, 31-342 Krak\'ow, Poland \\
krzysztof.wozniak@ifj.edu.pl} \\[0.2cm]
for the PHOBOS Collaboration \\
%
%
%
B.Alver$^4$,
B.B.Back$^1$,
M.D.Baker$^2$, M.Ballintijn$^4$,
D.S.Barton$^2$,
R.R.Betts$^6$,
A.A.Bickley$^7$,
R.Bindel$^7$,
W.Busza$^4$,
A.Carroll$^2$, Z.Chai$^2$, V.Chetluru$^6$,
M.P.Decowski$^4$,
E.Garc\'{\i}a$^6$,
N.George$^2$,
T.Gburek$^3$,
K.Gulbrandsen$^4$,
C.Halliwell$^6$,
J.Hamblen$^8$,
I.Harnarine$^6$,
M.Hauer$^2$,
C.Henderson$^4$,
D.J.Hofman$^6$,
R.S.Hollis$^6$,
R.Ho\l y\'{n}ski$^3$,
B.Holzman$^2$,
A.Iordanova$^6$,
E.Johnson$^8$,
J.L.Kane$^4$,
N.Khan$^8$,
P.Kulinich$^4$,
C.M.Kuo$^5$,
W.Li$^4$,
W.T.Lin$^5$,
C.Loizides$^4$,
S.Manly$^8$,
A.C.Mignerey$^7$,
R.Nouicer$^2$,
A.Olszewski$^3$,
R.Pak$^2$,
C.Reed$^4$,
E.Richardson$^7$,
C.Roland$^4$,
G.Roland$^4$,
J.Sagerer$^6$,
H.Seals$^2$,
I.Sedykh$^2$,
C.E.Smith$^6$,
M.A.Stankiewicz$^2$,
P.Steinberg$^2$,
G.S.F.Stephans$^4$,
A.Sukhanov$^2$,
A.Szostak$^2$,
M.B.Tonjes$^7$,
A.Trzupek$^3$,
C.Vale$^4$,
G.J.van~Nieuwenhuizen$^4$,
S.S.Vaurynovich$^4$,
R.Verdier$^4$,
G.I.Veres$^4$,
P.Walters$^8$,
E.Wenger$^4$,
D.Willhelm$^7$,
F.L.H.Wolfs$^8$,
B.Wosiek$^3$,
K.Wo\'{z}niak$^3$,
S.Wyngaardt$^2$,
B.Wys\l ouch$^4$ 
\\[0.2cm]
~$^1$~Argonne National Laboratory, Argonne, IL 60439-4843, USA \\
~$^2$~Brookhaven National Laboratory, Upton, NY 11973-5000, USA \\
~$^3$~Institute of Nuclear Physics PAN, Krak\'{o}w, Poland \\
~$^4$~Massachusetts Institute of Technology, Cambridge, MA 02139-4307, USA \\
~$^5$~National Central University, Chung-Li, Taiwan \\
~$^6$~University of Illinois at Chicago, Chicago, IL 60607-7059, USA \\
~$^7$~University of Maryland, College Park, MD 20742, USA \\
~$^8$~University of Rochester, Rochester, NY 14627, USA
\\[0.7cm]
\end{center}

\begin{center}
{\begin{minipage}{5truein}
                 \footnotesize
In the PHOBOS experiment, charged particles are measured
in almost the full solid angle. This enables the study of fluctuations and 
correlations in the particle production over a very wide kinematic range.
In this paper, we show results of a direct search for fluctuations identified by 
an unusual shape of the pseudorapidity distribution.  In addition, we use analysis of correlations of 
the multiplicity in similar pseudorapidity bins, placed symmetrically 
in the forward and backward hemispheres, to test the hypothesis of production
of particles in clusters.
\\[0.2cm]
\end{minipage}}\end{center}

\section{Introduction}
In the study of heavy ion collisions at the Relativistic Heavy Ion Collider (RHIC),
a new type of strongly interacting matter was discovered\cite{whitepaper}.
During a phase transition, rapid changes of the system's properties are possible and
unusual fluctuations may be observed\cite{qgpsignals}. 
Early predictions of signals of a quark-gluon plasma (QGP) suggested the
possibility of a large increase of total multiplicity or correlated
particle production in limited phase space. Although already the first
results from RHIC have shown only a moderate increase of charged particle 
multiplicity with collision energy, it is still possible 
that less apparent but significant fluctuations can be found.
A detailed analysis of fluctuations is also a source of information on
the particle production mechanism in the early stage of the collision and
in the later hadronization, both of which are important for understanding the properties of 
this new state of matter.

In our analysis, Au+Au collisions at $\sqrt{s_{_{\rm NN}}}$=200~GeV, 
the highest energy for heavy nuclei available at RHIC, are used.
In the first study, the most central (3\%) collisions are examined
in order to identify events with an unusual angular 
distribution of produced particles. The second study analyses 
multiplicity fluctuations
in forward and backward pseudorapidity bins and shows an enhancement with respect to 
purely statistical expectations. This may be explained by 
production of particles in clusters, which is modeled and compared to the experimental data. 

The PHOBOS detector\cite{detector} consists of several elements located 
mostly near the beam pipe.
Most important for the present analysis is the multiplicity detector, which 
has a uniquely large geometrical acceptance for measuring
produced particles. The hits from charged particles 
emitted at $|\eta|$~\textless~5.4 are registered 
in a single layer of silicon sensors, 
arranged as an octagonal tube
in the center of the apparatus (for $|\eta|$~\textless~3) and 
 mounted in three pairs of rings perpendicular to the beam direction 
for detection of particles emitted at smaller angles.
We measure the ionization energy deposited by the particle 
in the silicon and thus can estimate the number of particles
hitting the same pad by dividing the signal
by the value expected for a single particle. 
The multiplicity can be reconstructed using this estimation (``analog" method)
or by counting the hits and applying 
occupancy corrections (``hit counting" method).
The two methods
give consistent results and were successfully used to measure angular distribution
of charged particles
in the full acceptance of the PHOBOS detector\cite{multiplicity,dndeta}.
\section{Search for unusual events}
In the previous study\cite{qm2005rare} of the most central events (3\%), 
an excess of events with  
large multiplicities was found, whose rate was however correlated with
luminosity. The fraction of events with a $dN/d\eta$ distribution deviating 
from the mean shape also increased with the event rate. 
The most probable explanation of such results is the registration of two Au+Au
collisions as one event (pileup). 
In this paper, we apply a more efficient rejection of pileup events
and focus the analysis on the  $dN/d\eta$ distribution. 
For each of over 1,900,000
most central events we determine the raw number of particles
as a function of pseudorapidity, $M(\eta)$, using the ``hit counting" method
without occupancy corrections. We also calculate
the mean value $\langle M(\eta) \rangle$ and variance $\sigma^{2}_{_{\rm M}}(\eta)$
in narrow bins of vertex position (to minimize acceptance fluctuations).
The deviation of the $\eta$ distribution for an event is then measured by:
\begin{equation}
\chi^2_{_{\rm NDF}} = \frac{1}{N_{\eta \rm bins}}  \sum_{\eta}\frac{(M(\eta) - S \langle M(\eta)\rangle)^{2}}{\sigma^{2}_{_{\rm M}}(\eta)}  \label{chi2def}
\end{equation}
where $N_{\eta \rm bins}$ denotes the number of bins in $\eta$ in which $M(\eta)$ is defined
and $S$ is the fitted scaling factor accounting for varying total multiplicity.
We define as unusual the events with $\chi^2_{_{\rm NDF}}$\textgreater 3,
a value which should not be exceeded in our sample.
\begin{figure}[tb]
\centerline{
\epsfxsize=7cm\epsfbox{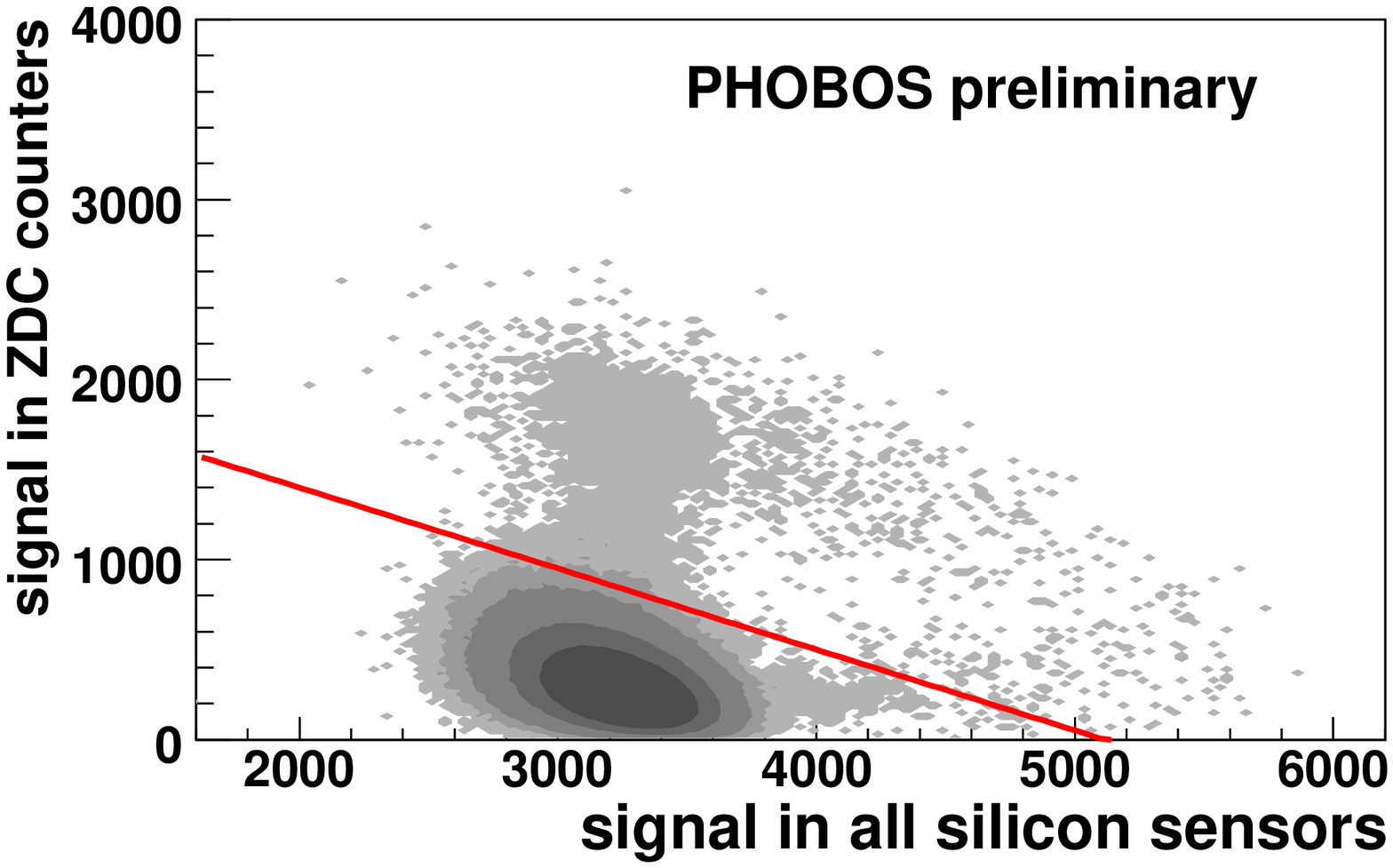}
 \hspace{0.5cm}
 \epsfxsize=7cm\epsfbox{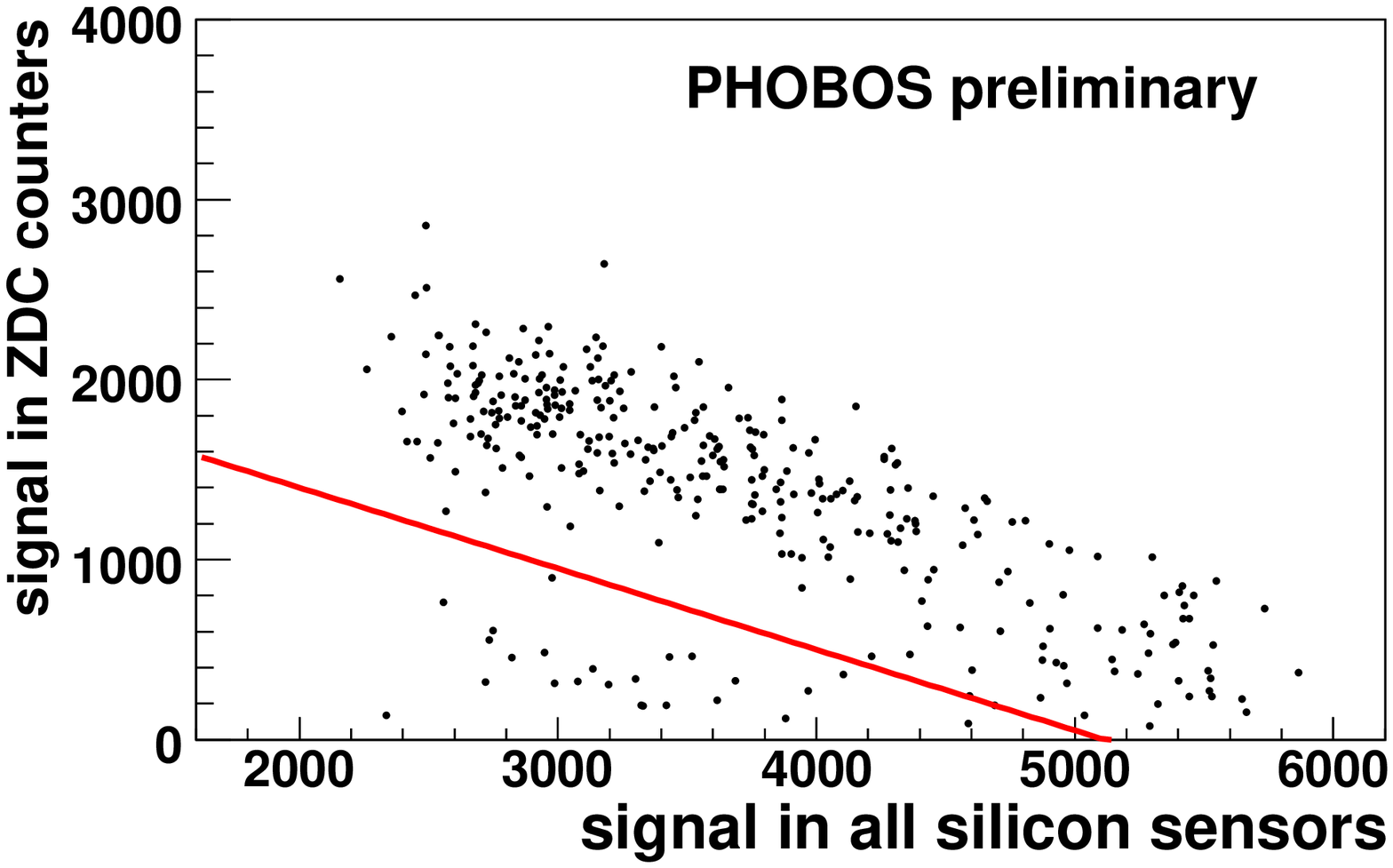}
}
\caption{Rejection of pileup events with large signal in the ZDC counters (above the line)
for all central events (left) and events with unusual $dN/d\eta$ shape (right).}
\end{figure}
In the previous study, 
pileup events were identified by comparison of signals in the silicon sensors (long
signal integration time) with signals in the scintillator trigger counters
(short signal integration time). This allows to eliminate merging of Au+Au collisions
from different bunch crossings. In this paper, we also compare signals in the silicon
sensors with the signal left by spectator neutrons in the Zero Degree Calorimeters (ZDC). 
For the most central Au+Au collisions, the ZDC signal should be small, thus the events
above the lines in Fig.~1 are rejected, as most probably the central collision
is accompanied by a peripheral one.
This cut removes over 90\% of the events with $\chi^2_{_{\rm NDF}}$\textgreater 3 and, together with
the previous cut, reduces the tail of large $\chi^2_{_{\rm NDF}}$ significantly (Fig.~2).
Properties of the remaining unusual events will be investigated in a future study, 
but we know already that the fraction of central events with large 
deviations of their $dN/d\eta$ shape is smaller than 10$^{-5}$.
\begin{figure}[th]
\centerline{
 \epsfxsize=7cm\epsfbox{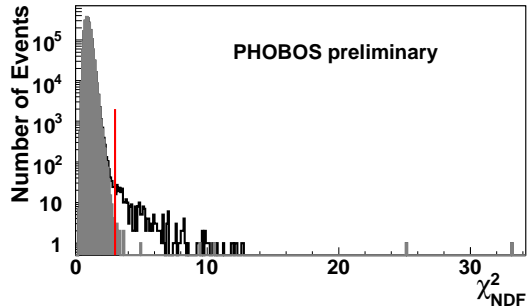}
}
\caption{Distribution of $\chi^2_{_{\rm NDF}}$ for all central events (histogram) and these accepted 
by the pileup cuts (shaded area).}
\end{figure}
\section{Forward-backward fluctuations}
Another aspect of multiplicity fluctuations 
is studied by comparing 
the multiplicity measured in the forward and backward hemispheres, reconstructed
event-by-event in pseudorapidity bins of varying size and position. 
The multiplicity $N_{_{\rm F}}$ reconstructed in a bin in the forward $\eta$ region is compared
to the multiplicity $N_{_{\rm B}}$ in a similar bin placed symmetrically at negative $\eta$
using a correlation parameter $C$: 
\begin{equation}
 C = \frac{N_{_{\rm F}} - N_{_{\rm B}}}{ \sqrt{N_{_{\rm F}}+N_{_{\rm B}}} }  \label{cdef}
\end{equation}
which is calculated as a function of position of the center of 
the positive $\eta$ bin or of its width, $\Delta\eta$. 
In the comparison of bins covering similar kinematical regions the expected mean
value of $C$ is zero. If, in addition, multiplicity fluctuates only statistically
according to the binomial partitioning of $N_{_{\rm F}}+N_{_{\rm B}}$, we obtain 
the variance $\sigma^{2}(C)=1$. Values of $\sigma^{2}(C)$ different from 1 would indicate 
the presence of nonstatistical fluctuations, possibly due to correlations 
between primary particles.

In this analysis, a simplified version of the ``analog" method based on 
the energy deposited by charged particles in the octagonal multiplicity detector
is used to determine $N_{_{\rm F}}$ and $N_{_{\rm B}}$.
Using detector simulations,
the value of $\sigma^2(C)$ is corrected for modifications due to secondary production,
Landau energy loss fluctuations, and acceptance effects\cite{fbc}.
The reconstructed values of $\sigma^{2}(C)$ are larger than 1, as  
is expected in the case of particle production in two steps.
For a simple example, we can assume that initially produced objects (clusters)
decay into  $k$ particles, all of which are registered 
in the same $\eta$ bin. If the primary clusters are produced
randomly (giving $\sigma^{2}(C_{\rm cluster})$=1), the variance of $C$ for final
particles grows to  $\sigma^{2}(C)=k$. 
If the distribution of cluster size, $k$, has a mean value of $\langle k\rangle$ 
and variance  $\sigma_{\rm k}$, the relevant value is 
$k_{\rm eff}=\langle k\rangle+\sigma^2_{\rm k}/\langle k\rangle$.
In a more realistic case, particles originating from the same cluster
are distributed in pseudorapidity, and thus usually are not all contained 
in the same $\eta$ bin used in the calculation of $C$. In addition, 
if the forward and backward
bins are close, particles from a single cluster may be emitted into opposite bins. 
These two effects 
decrease the value of  $\sigma^{2}(C)$ in a bin-width and bin-position dependent way. 

\section{Simple Cluster Models}
The hypothesis of cluster production in Au+Au collisions can be tested 
using simple models generating clusters which then decay into particles.
Any such model should reproduce the general properties of Au+Au events:
the experimental pseudorapidity distribution\cite{dndeta} 
and the transverse momentum distribution\cite{ptfit},
However, perfect agreement is not necessary, as we have found 
that modifications of the $dN/d\eta$ shape 
or the $\langle p_{\rm t}\rangle$ value cause only small changes of the impact
of clusters.

The Isotropic Cluster Model (ICM) assumes that clusters have negligible
transverse momentum and decay into particles emitted isotropically 
in the clusters' rest frame. In this frame, the pseudorapidity distribution of particles
has a shape similar to a Gaussian function, but with longer tails.
The $\sigma(\eta)$ of isotropic clusters is about 1.0, and is moderately modified
by the Lorentz transformation of the final particles to the laboratory frame, 
especially for the largest $|\eta|$ values (it is about 15-20\% smaller at $|\eta|$=3).
In this model, the multiplicity of a cluster can be freely selected.
In order to obtain a continuous spectrum of $k_{\rm eff}$ values, 
we generated $k$ from a Poisson distribution with an appropriate mean value.

Exactly isotropic decay of each cluster is an extreme case, in which
$\sigma(\eta)$ is the largest. Such angular symmetry is not preserved
in decays of resonances, which are the most
natural candidates for clusters.
The momentum of an emitted particle, for example a pion 
in a decay $R(M_{\rm 1})~\longrightarrow~R(M_{\rm 2}) + \pi$, depends mostly on the mass difference
$\Delta M = M_{\rm 1}-M_{\rm 2}$. Only the values $\Delta M <$~650~MeV 
lead to momenta of pions
smaller or consistent with the experimental mean $p_{\rm t}$.
In our Resonances Cascade Model (RCM) we select a chain of decays starting from relatively
light particles and thus small $\Delta M$ in decays:
\begin{equation}
\Delta(1900)~~\longrightarrow~~\Delta(1600) + \pi~~\longrightarrow~~\Delta(1232) + 2\pi~~\longrightarrow~~N + 3\pi    \label{decaydef}
\end{equation}
In the later analysis we assume that one of the final particles 
is neutral, thus the size of the cluster is always 3. 
In order to reproduce even approximately the experimental $p_{\rm t}$
distribution of pions and protons, the transverse momentum of the cluster
($\Delta(1900)$ in this case) has to be of the order of 2~GeV/c. 
Such large initial $p_{\rm t}$ causes the final particles to be close in $\eta$
and $\sigma(\eta)$ is only about 0.6. For other resonance cascades with larger
values of $\Delta M$ in intermediate decays, the initial $p_{\rm t,cluster}$ 
is smaller and thus $\sigma(\eta)$ is larger.

The two cluster models were used to generate several samples of 100,000 events
containing particles from many clusters. The multiplicity of generated charged particles
was counted in appropriate $\eta$ bins and values of $\sigma^{2}(C)$
were calculated. These are compared with the data from the analysis\cite{fbc} 
of Au+Au collisions at $\sqrt{s_{_{\rm NN}}}$~=~200~GeV.
In Fig.~3 (left plot) we show the dependence of $\sigma^{2}(C)$ on $\Delta\eta$, 
the width of the $\eta$ bins. The ICM results are presented for
two different values of effective cluster size, $k_{\rm eff}$=3.4 and $k_{\rm eff}$=4.3,
which fit the experimental results for central (0-20\%) and 
semi-peripheral (40-60\%) Au+Au collisions, respectively. The same cluster sizes 
are then used to obtain ICM predictions for the dependence of  $\sigma^{2}(C)$ 
on the position in $\eta$ of the bins, all with a width of 0.5 (Fig.~3, right plot). 
In the case of the RCM,
only one value of $k_{\rm eff}$=3 is used and results are presented in Fig.~3.
The dependence of $\sigma^{2}(C)$ on $\Delta\eta$ shows that in peripheral
collisions the effective cluster size is larger by about one. 
This $k_{\rm eff}$ difference may be smaller if the clusters in central collisions
are narrower in $\eta$, as the dependence of $\sigma^{2}(C)$ on $\eta$ suggests. 
The two models (but especially ICM) described in this study represent the extreme cases of
the narrowest and the widest possible pseudorapidity clusters.  The results allow us
to set lower (from RCM) and upper (from ICM) limits for the cluster size 
(according to Fig.~3, left plot, they are $\sim$2.5 and $\sim$4.3 respectively).
In a more realistic model, a mixture of clusters with different widths and 
sizes should be used, leading to $k_{\rm eff}$ values between the RCM and ICM predictions.
However, the data from forward-backward correlations 
are not precise enough to determine $k_{\rm eff}$
and cluster $\eta$-width at the same time.

\begin{figure}[tb]
\centerline{
 \epsfxsize=7cm\epsfbox{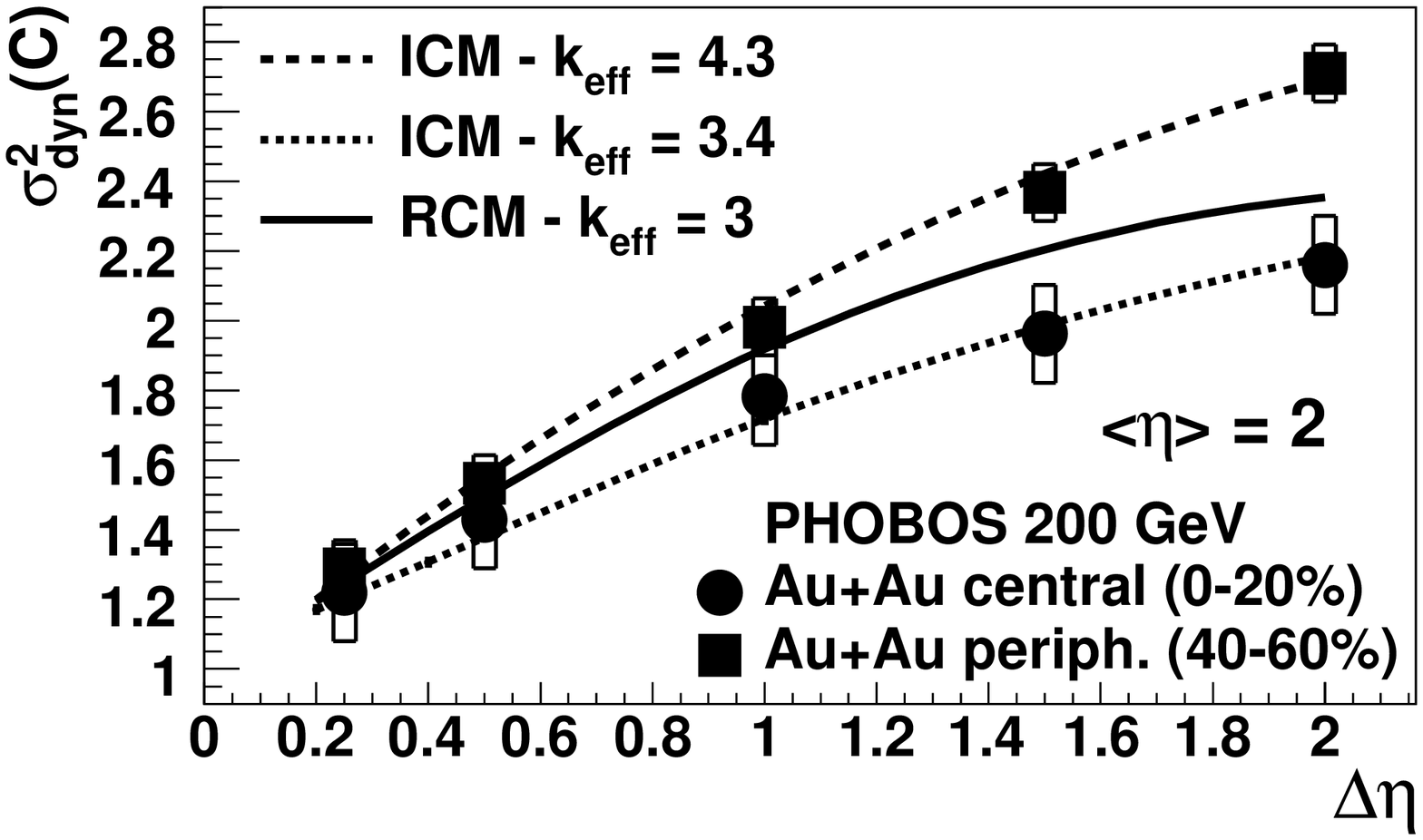}
 \hspace{0.5cm}
 \epsfxsize=7cm\epsfbox{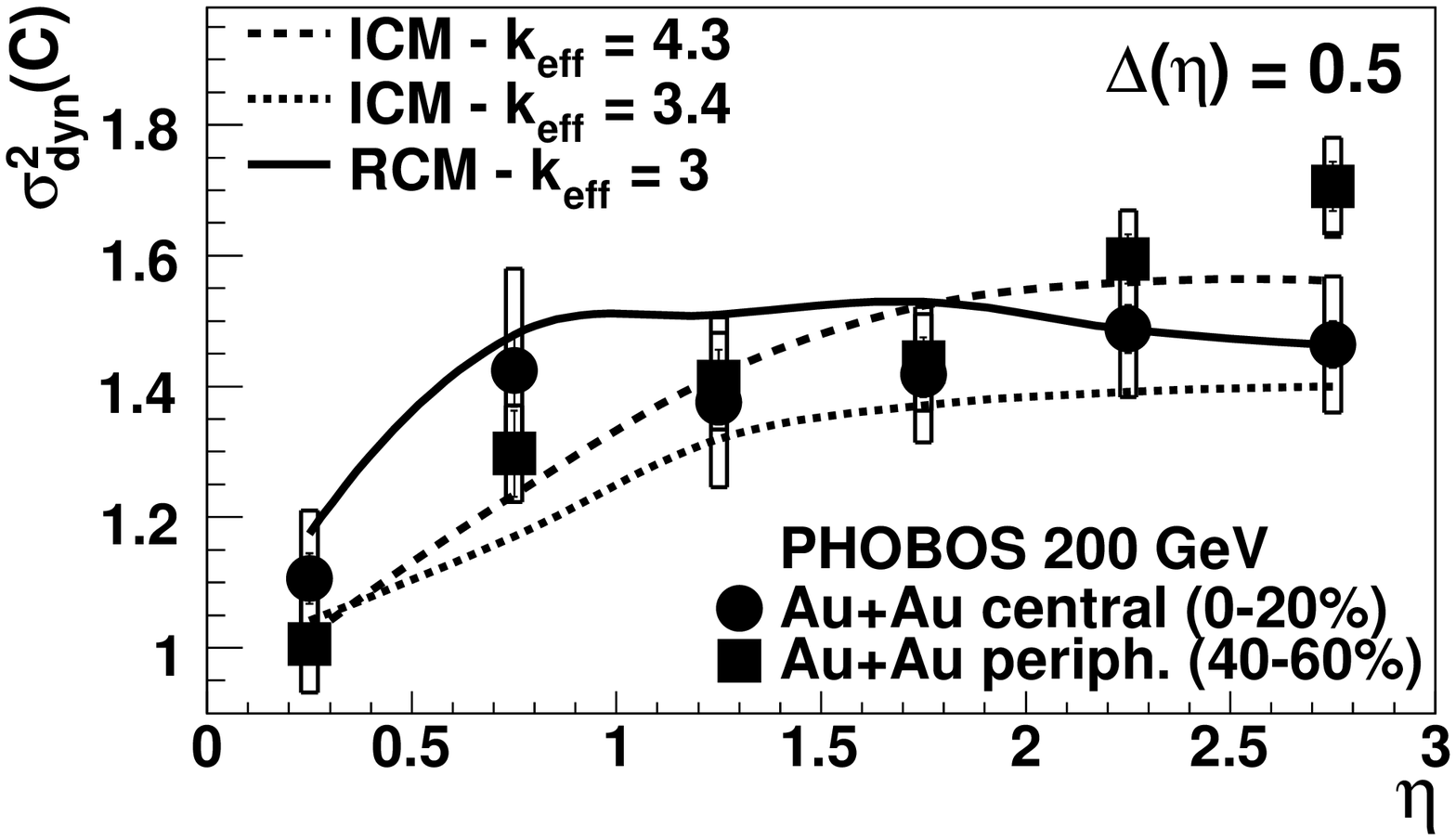}
}
\caption{Comparison of reconstructed $\sigma^2(C)$ with the predictions from
two cluster models: the dependence on the width of the $\eta$ bin (left)
and on the position of the bin center (right).}
\end{figure}

\section{Conclusions}
A large sample of the most central 3\% of Au+Au collisions at $\sqrt{s_{_{\rm NN}}}$~=~200 GeV
was examined in the search for events with an unusual shape of the $dN/d\eta$ distribution.
After pileup removal only a few such events were left, which allows
us to set an upper limit for the probability of such fluctuations at 10$^{-5}$.

The forward-backward multiplicity correlations are sensitive
to multiplicity fluctuations at short distances in $\eta$,
which are expected if particles are produced in clusters.
Using two simple clusters models, we found that the 
effective cluster size consistent with the experimental data
is between 2.5 and 4.

\section*{Acknowledgements}
{\small This work was partially supported by U.S. DOE grants 
DE-AC02-98CH10886,
DE-FG02-93ER40802, 
DE-FG02-94ER40818,  
DE-FG02-94ER40865, 
DE-FG02-99ER41099, and
DE-AC02-06CH11357, by U.S. 
NSF grants 9603486, 
0072204,            
and 0245011,        
by Polish KBN grant 1-P03B-062-27(2004-2007),
by NSC of Taiwan Contract NSC 89-2112-M-008-024, and
by Hungarian OTKA grant (F 049823).
}


\noindent

\begin{thebibliography}{0}
\bibitem{whitepaper} B.B. Back {\it et al.}, Nucl. Phys. A 757 (2005) 28.
\bibitem{qgpsignals} J.W. Harris and B. M\"{u}ller, Ann. Rev. Nucl. Part. Sci., vol 46 (1996) 71.
\bibitem{detector} B.B. Back {\it et al.}, Nucl. Instr. and Meth. A 499 (2003) 603.
\bibitem{multiplicity} B.B. Back {\it et al.}, Phys. Rev. Lett. 87 (2001) 102303.
\bibitem{dndeta} B.B. Back {\it et al.}, Phys. Rev. Lett. 91 (2003) 052303.
\bibitem{qm2005rare} G.S.F.Stephans {\it et al.}, Nucl. Phys. A 774 (2006) 639.
\bibitem{fbc} B.B. Back {\it et al.}, Phys. Rev. C74 (2006) 011901(R).
\bibitem{ptfit} B.B. Back {\it et al.}, Phys. Rev. C 70 (2004) 051901(R).
\end{thebibliography}
\end{document}